\documentclass[aps,prd,preprint,nofootinbib]{revtex4}
\usepackage{amsfonts}
\usepackage{mathrsfs}
\usepackage{graphicx}
\usepackage{amsmath}
\usepackage{amssymb}
\usepackage{bm,multirow}
\usepackage{multirow}
\usepackage{epsfig}
\usepackage{graphicx}
\newcommand{\bqa}{\begin{eqnarray}}
\newcommand{\eqa}{\end{eqnarray}}
\newcommand{\beq}{\begin{equation}}
\newcommand{\eeq}{\end{equation}}
\graphicspath{{fig/}{dia/}} \DeclareGraphicsExtensions{.eps}
\usepackage{subfig}
\begin{document}
\baselineskip 20pt
\title{Two-loop QCD Corrections to $B_c$ Meson Leptonic Decays}


\author{Long-Bin Chen$^1$\footnote{chenglogbin10@mails.ucas.ac.cn}}

\author{Cong-Feng Qiao$^{1,2}$\footnote{qiaocf@ucas.ac.cn, corresponding author}}

\affiliation{$^1$School of Physics, University of Chinese Academy of Sciences, YuQuan Road 19A, Beijing 100049, China\\
$^2$CAS Center for Excellence in Particle Physics, Beijing 100049, China}

\author{~\\}

\begin{abstract}
The two-loop QCD radiative corrections to the $B_c$ meson leptonic decay
rate are calculated in the framework of NRQCD factorization formalism. Two types of master integrals appearing in the calculation are obtained analytically for the first time. We get the short-distance coefficient of the leading matrix element to order $\alpha_s^2$ by matching the full
perturbative QCD calculation results to the corresponding NRQCD ones. The result in this work helps the evaluation of the $B_c$ leptonic decay constant, as well as the Cabibbo-Kobayashi-Maskawa matrix element $|V_{cb}|$, to the full next-to-next-to-leading order degree of accuracy.

\vspace {7mm} \noindent {PACS number(s): 12.38.Bx, 13.25.Gv,
14.40.Be }

\end{abstract}
\maketitle


The advent of non-relativistic Quantum Chromodynamics (NRQCD) factorization formalism causes investigations on heavy quarkonium more reliable \cite{NRQCD}, which improves the understanding of strong interaction. It has been noted that for quarkonium production and decays, in many cases the leading order calculation in the framework of NRQCD is inadequate. And, mostly the discrepancy between leading order calculation and experimental result can be rectified by including higher order corrections, which has
stimulated various investigations in this respect.

$B_c$ meson system, which contains two different heavy quark flavors, has some peculiar natures different from heavy quarkonium, and recently attracts great interest, especially with the progress of the LHCb experiment \cite{review-lhcb}. Though $B_c$ meson is very elusive in experiment, the feedback from the investigation on it is extremely great, e.g. on some aspects of quantum chromodynamics(QCD), weak interaction and even new physics. Of the ${\bar b} c$ system, the higher excited states will mostly cascade down to the ground state, the pseudoscalar $B_c$ meson, through hadronic or electromagnetic transitions, which then decays to lighter hadrons or leptons via weak interaction.

By virtue of nonrelativistic QCD(NRQCD) formalism, the $B_c$ meson decay amplitude may be expressed as perturbative QCD(pQCD) calculable short-distance coefficients multiplied by non-pertubative NRQCD
matrix elements. The expression for leading order(LO) $B_c$ meson leptonic decay width is simple and known for long, and the next-to-leading order calculation was completed by Braaten and Fleming two decades ago \cite{bc1loop}. In this work, we compute in pQCD the two-loop radiative corrections to the pseudoscalar $B_c$ meson leptonic decay rate. i.e., the short-distance coefficient for the leading matrix element at the next-to-next-to-leading order(NNLO), by matching the perturbative result in full QCD with the corresponding perturbative calculation in NRQCD.

The calculation of massive two-loop Feynman integrals is somehow tough, especially with two mass scales. For this reason, there are only a few master integrals with different massive propagators have been accomplished \cite{remid,huber1,huber2}. The method of differential equations turns out to be an efficient and powerful technique for the calculation of Feynman integrals. Recently, it was found by Henn that the solution of differential equation will be simplified considerably if the bases of master integrals are chosen properly \cite{newde}. In our calculation, by employing the technique of differential equation and choosing certain bases of master integrals, we successfully obtain the master integrals required in the calculation of two-loop QCD corrections to $B_c$ leptonic decays.


The $B_c$ meson leptonic decays, $B_c \rightarrow \l\ \nu_l$ with $l$ being e, $\mu$, or $\tau$, are heavy-quark-annihilation processes through axial-vector current, which are very important to the study of $B_c$ physics while have not been, but expected to be, measured. Theoretically the decay rates can be formulated as:
\begin{equation}
\Gamma(B_c \to \ell^+ \nu_\ell) \;=\;
{1 \over 8 \pi} |V_{bc}|^2 G_F^2 M f_{B_c}^2 m_\ell^2
	\left( 1 - {m_\ell^2 \over M^2} \right)^2 ,
\label{width}
\end{equation}
where $V_{bc}$ denotes the CKM matrix element; $M$ and $m_\ell$ stand for masses of $B_c$ meson and charged leptons, respectively; and $G_F$ is the Fermi coupling constant of weak interaction. Generally, the $B_c$ decay constant $f_{B_c}$ is defined through the transition matrix element of charged weak current, as
\begin{equation}
\langle 0 | {\bar b} \gamma^\mu \gamma_5 c | B_c(p) \rangle
\; = \; i f_{B_c} p^\mu\ ,
\label{fbc}
\end{equation}
which parameterizes the strong interaction effects and contains both perturbative and nonperturbative contributions.

The short-distance contribution can be isolated and calculated in perturbation theory, by matching the charged weak current in QCD to a series of operators in NRQCD. In the rest frame of $B_c$ system, up to corrections of order $v^4$, the relative velocity of heavy quarks within the meson, the matching relation reads \cite{bc1loop}
\begin{equation}
\langle 0 |{\bar b} \gamma^0 \gamma_5 c | B_c(p) \rangle
\;=\; C_0 \; \langle 0 |\chi_b^\dagger \psi_c | B_c(p) \rangle
\;+\; C_2 \; \langle 0 | ({\bf D} \chi_b)^\dagger \cdot {\bf D} \psi_c | B_c(p) \rangle
\;+\; \ldots\ ,
\label{opex}
\end{equation}
where $C_0$ and $C_2$ are short-distance coefficients that depend on the heavy quark masses, renormalization scale $\mu$, and strong interaction coupling $\alpha_s$. The coefficients $C_0$ and $C_2$ will be determined by matching the perturbative calculation of the matrix elements in full QCD with what obtained in the framework of NRQCD \cite{bc1loop}. The coefficient $C_0$ was obtained at one-loop order in Ref.\cite{bc1loop}, i.e.,
\begin{equation}
C_0 \;=\;
1 \;+\; {\alpha_s \over \pi}
	\left[{m_b-m_c \over m_b+m_c} \log {m_b \over m_c}
		- 2 \right] \;+\; ({\alpha_s \over \pi})^2 co_2(m_c,m_b,\mu)\ .
\label{C0}
\end{equation}
This work is about to calculate the NNLO QCD corrections to the short-distance coefficient of the leading order matrix element in $v^2$ expansion analytically, i.e. $co_2$, by which the theoretical prediction for $B_c$ leptonic decay rates will come up to the NNLO accuracy.

\begin{figure}[h,t,m,u]
\begin{center}
\includegraphics[scale=0.3]{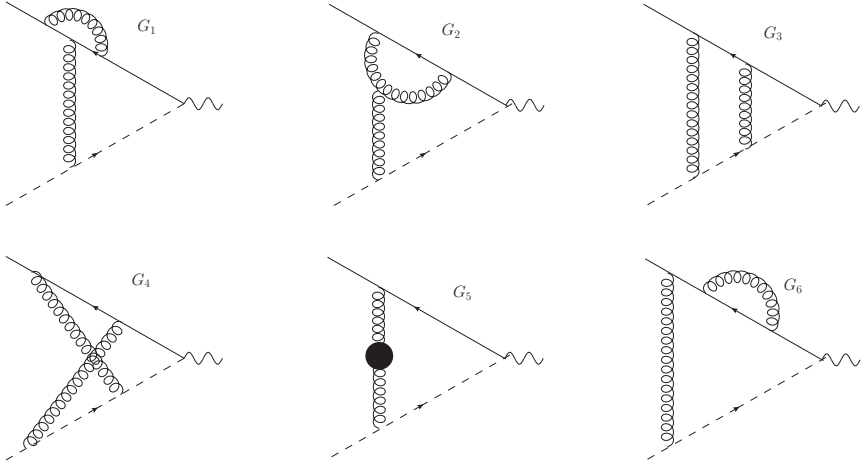}
\caption{Two loop diagrams that contribute to the concerned processes. For $G_{1,2,6}$, the symmetric diagrams are implied. The diagram $G_5$ contains contributions from massless fermions, gluons, ghosts, and the massive fermions. \label{fig1}}
\end{center}
\end{figure}


In calculating the two-loop contributions in full QCD, we first revisit the one-loop QCD corrections to $B_c$ meson leptonic decays.
At one-loop order, there is only one diagram in Feynman gauge, and the Mathematica package {\bf FeynArts} \cite{feynarts} is employed to generate the amplitude. {\bf FeynCalc} \cite{Feyncalc} combined with code written by ourselves are used to manipulate the $\gamma$-matrix algebra and spin projections, and {\bf FIRE} \cite{fire} together with {\bf \$Apart} \cite{apart} are employed to reduce all the related integrals into a set of master integrals. In one-loop case, the integrals are merely subject to two massive tadpoles. After performing the standard renormalization procedure, we then obtain the coefficient $C_0$ at the order of $\alpha_s$, which agrees with Ref. \cite{bc1loop}.

The topologically independent two-loop order Feynman diagrams are schemetically shown in Fig.1, where the solid line represents for bottom quark and dashed line for charm quark. Note, within the figure, the solid-dashed lines exchanged diagrams of $G_{1,2,6}$ are implied. Though the $W^+$ boson may couple to charm and bottom quarks through both vector and axial-vector currents, in practice only the axial-vector current contributes to the annihilation decays of pseudoscalar $B_c$. The calculation of two-loop amplitude takes the same procedure as in one-loop case. The hardest part of the two-loop calculation in this work resides in the evaluation of the master-integrals, as explained in the following.


\begin{figure}[h,t,m,u]
\begin{center}
\includegraphics[scale=0.5]{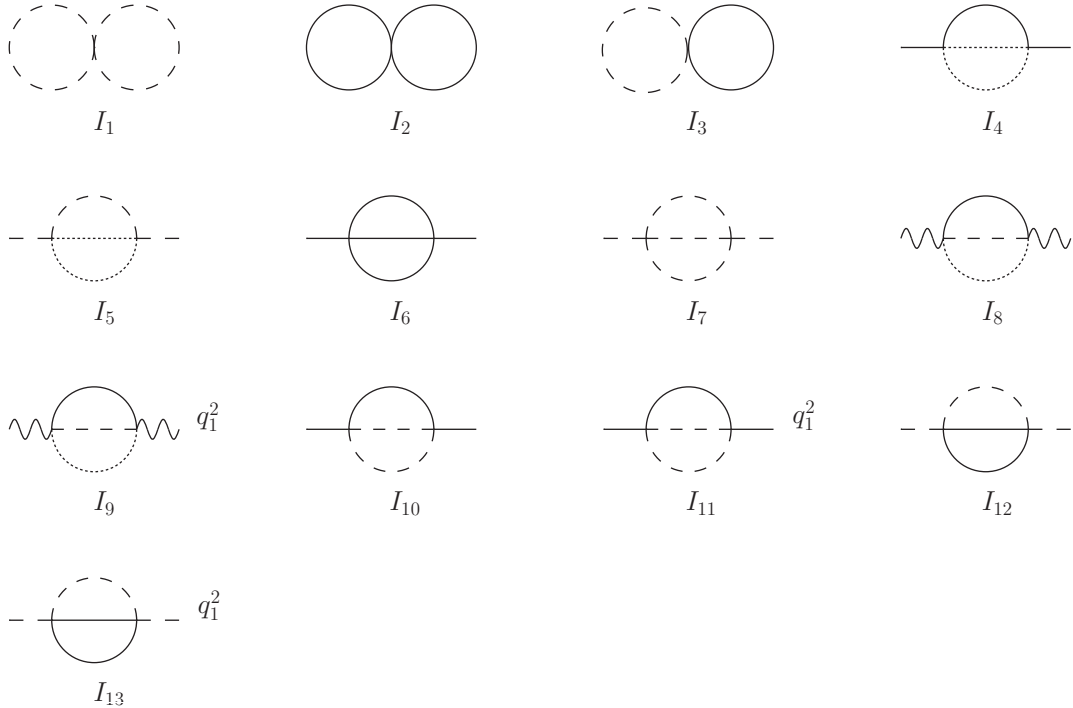}
\caption{Two-loop diagrams of various master integrals. \label{fig2}}
\end{center}
\end{figure}

The master integrals we confront in the calculation are shown in Fig. 2, where dotted, dashed, solid lines correspond respectively to the gluon, charm quark and bottom quark propagators, the incoming and outgoing wavy lines of diagrams $I_8$ and $I_9$ have a fictitious invariant mass of $p^2=(m_c+m_b)^2$. In order to obtain the master integrals by employing the technique of differential equation and for the sake of compactness, we attribute the two heavy quark masses to a new parameter $x=\frac{m_c}{m_b}$, and normalize the loop integrations in terms of
\begin{equation}
\int[
d^d q] = \frac{e^{\gamma_E \epsilon}}{i\pi^{D/2}} \left( \frac{\mu^2}{m_b^2}
\right)^{-\epsilon} \mu^{2\epsilon}\int d^d q \, ,
\label{measure}
\end{equation}
with $\gamma_E$ the Euler constant and $\mu$ the renomalization scale.

The integrals $I_1, I_2, I_3$ are just the multiplication of two tadpoles, and the integration for tadpole diagram can be found for instance in \cite{gbell}. The $I_4$ type of integral can be obtained in this procedure: first use {\bf AMBRE}-package \cite{ambre} to transform the integral to Mellin-Barnes representation, then use {\bf MB}-package \cite{mb} to single out the poles in $\epsilon$ expansion, and last evaluate the integral by closing a proper contour and summing up residues. In the end, we find
\begin{equation}
I_4 =\frac{1}{2\epsilon^2}+\frac{5}{4\epsilon}+\big(\frac{11}{8}+\frac{5\pi^2} {12}\big)+\frac{\epsilon}{48}\big(-165+50\pi^2+176 \zeta(3)\big) \, .
\label{measure}
\end{equation}
Except for changing $m_b$ to $m_c$ in the normalization (\ref{measure}), the integral $I_5$ is just the same as $I_4$, which agrees with Ref. \cite{gbell}.

$I_6$ can proceed in a similar way as $I_4$, while with the coefficient of $\frac{1}{(D-4)^2}$ after the reduction. Up to the $\epsilon^2$, the $I_6$ reads:
\begin{eqnarray}
I_6 &=&\frac{3}{2\epsilon^2}+\frac{17}{4\epsilon}+\big(\frac{59}{8}+ \frac{\pi^2}{4}\big)+\epsilon\big(\frac{65}{16}+\frac{49\pi^2}{24}- \zeta(3)\big)+\nonumber \\
& &\frac{\epsilon^2}{480}\big(-16755+4750 \pi^2 + 14 \pi^4 - 3840 \pi^2 \ln(2) + 12080 \zeta(3)\big) \, .
\label{measure}
\end{eqnarray}
Then integral $I_7$ can be readily obtained the same way as $I_5$. Note that the integral $I_6$ will serve as the boundary condition for integrals $I_{10-13}$.

The master integrals for $I_8$ to $I_{13}$ are
\begin{eqnarray}
I_8 &=& \int[d^d q_1][d^d q_2] \frac{1}{q_2^2(q_1^2-2p_b\cdot q_1)((q_2-q_1)^2-2x (q_2-q_1)\cdot p_b)}\ ,\\
I_9 &=& \int[d^d q_1][d^d q_2] \frac{q_1^2}{q_2^2(q_1^2-2 p_b\cdot q_1)((q_2-q_1)^2-2 x (q_2-q_1)\cdot p_b)m_b^2}\ , \\
I_{10} &=& \int[d^d q_1][d^d q_2] \frac{1}{(q_1^2-2p_b\cdot q_1)(q_2^2-2 x p_b\cdot q_2)((q_2-q_1)^2-2x (q_2-q_1)\cdot p_b)}\ , \\
I_{11} &=& \int[d^d q_1][d^d q_2] \frac{q_1^2}{(q_1^2-2p_b\cdot q_1)(q_2^2-2 x p_b\cdot q_2)((q_2-q_1)^2-2x (q_2-q_1)\cdot p_b)m_b^2}\ ,~~~~~ \\
I_{12} &=& \int[d^d q_1][d^d q_2] \frac{1}{(q_1^2+2p_b\cdot q_1)(q_2^2-2 x p_b\cdot q_2)((q_2+q_1)^2+2 (q_2+q_1)\cdot p_b)}\ , \\
I_{13} &=& \int[d^d q_1][d^d q_2] \frac{q_1^2}{(q_1^2+2p_b\cdot q_1)(q_2^2-2 x p_b\cdot q_2)((q_2+q_1)^2+2 (q_2+q_1)\cdot p_b)m_b^2}\ ,
\end{eqnarray}
which satisfy differential equations in $x$.

In deriving the differential equations for $I_{8-13}$, the {\bf FIRE} package was employed. We find the following differential equations exist:
\begin{eqnarray}
\label{diff-I9}
\frac{d I_8}{d x} &=& \frac{4D-10+(5D-14)x-x^2}{x(1+x)(2+x)}I_8 +\frac{6-3D}{2x(2+x)}I_9+\frac{D-2}{2x(1+x)(2+x)}I_3\ ,\\
\frac{d I_9}{d x} &=& \frac{2x(7-3D)}{(1+x)(2+x)}I_8 +\frac{D-4+x(2D-5)}{(1+x)(2+x)}I_9+\frac{2(D-2)}{x(1+x)(2+x)}I_3\ .
\end{eqnarray}
In Ref. \cite{newde} the author suggested that properly chose of master integrals can lead to significant simplifications of the differential equations. To employ this technique, we need to find a pair of bases $g_8$ and $g_9$, by which the above differential equations can be transformed into the following form:
\begin{eqnarray}
\partial_x g(\epsilon,x)=\epsilon A(x)g(\epsilon,x) + \phi(\epsilon,x)\ .
\label{norm}
\end{eqnarray}
Here, $\phi(\epsilon,x)$ is a known function and can be expressed as harmonic polylogarithms of argument $x$. Note, the differential equation (\ref{norm}) contains poles in $x=0,-1,-2$. Suppose $g_8=b_8(x) I_8 + b_9(x) I_9$, with $b_8(x)$ and $b_9(x)$ being rational functions of $x$ and can be expressed as:
\begin{eqnarray}
b_8(x)=\sum_{i=-n_0}^{n_1}\frac{y_{1i}}{x^i}+\sum_{i=1}^{n_2} \frac{y_{2i}}{(1+x)^i}+\sum_{i=1}^{n_3}\frac{y_{3i}} {(2+x)^i}\ , \\
b_9(x)=\sum_{i=-n_0}^{n_1}\frac{z_{1i}}{x^i}+\sum_{i=1}^{n_2}\frac{z_{2i}} {(1+x)^i}+\sum_{i=1}^{n_3}\frac{z_{3i}}{(2+x)^i}\ ,
\end{eqnarray}
by trial and error, we finally sort out two bases that satisfy the canonical form of Eq.(\ref{norm}), i.e.
\begin{eqnarray}
g_8 &=& \frac{x^2}{1+x}I_8+I_9\ , \\
g_9 &=& \frac{2+6x+5x^2}{2x^3(1+x)(2+x)}I_8-\frac{2x+1}{x^3(2+x)}I_9\ .
\end{eqnarray}
Notice the boundary station at $x=1$ can be solve the same way as $I_4$ and $I_6$, then the integrals can be solved iteratively in terms of logarithms and polylogarithms. We use {\bf HPL}-packages \cite{hpl} to transform the logarithms and polylogarithms into harmonic polylogarithms with simple argument. In the end, the master integral $I_8$ obtained as
\begin{eqnarray}
I_8 &=& \frac{1+x^2}{2\epsilon^2}+ \frac{1}{4\epsilon}\big(5-2x+5x^2-8x^2H_0(x)\big)+ \frac{1}{8}(11-26x+11x^2)\nonumber\\
& &+\ \frac{\pi^2}{12(1+x)^2}\big(5+10x+2x^2+18x^3+9x^4 \big)-\frac{x(-2+5x+5x^2)H_0(x)}{1+x}\nonumber\\
& &+\ \frac{2x^2(2+6x+3x^2)H_{0,0}(x)}{(1+x)^2}+ (1-x^2)(H_{-,0}(x)-H_{+,0}(x)) \nonumber\\
& &+\ \epsilon\big[\frac{-5(11+46x+11x^2)}{16}+ \frac{(25+8x-66x^2+56x^3+45x^4) \pi^2}{24(1+x)^2}\nonumber\\
& &-\ \frac{x(-78+(2\pi^2-45)x+22(2\pi^2+3)x^2+11(2\pi^2+3)x^3) H_0(x)}{6(1+x)^2}\nonumber\\
& &+\ \frac{(2+4x+2x^3+x^4)\pi^2}{3(1+x)^2}(H_+(x)-H_-(x))+ \frac{x(-8+28x^2+15x^3)H_{0,0}(x)}{(1+x)^2} \nonumber\\
& &-\ \frac{5+7x-7x^2-5x^3}{2(1+x)}\big(H_{+,0}(x)-H_{-,0}(x) \big)-\frac{4x^2(2+10x+5x^2)}{(1+x)^2}H_{0,0,0}(x) \nonumber\\
& &+\ \frac{2x^3(2+x)}{(1+x)^2}\big(H_{0,-,0}(x)-H_{0,+,0}(x) \big)+\frac{2(2+4x-2x^3-x^4)}{(1+x)^2} \big(H_{+,0,0}(x) \nonumber\\
& &-\ H_{-,0,0}(x)\big)+(1-x^2)\big(H_{+,+,0}(x)+ H_{-,-,0}(x)-H_{+,-,0}(x)-H_{-,+,0}(x)\big)\nonumber\\
& &+\ \frac{11+22x-2x^2+46x^3+ 23x^4}{3(1+x)^2}\zeta(3)\big]\ .
\label{master-I8}
\end{eqnarray}
With (\ref{master-I8}), $I_9$ can be derived out from equation (\ref{diff-I9}) directly.

For integrals $I_{10}$ and $I_{11}$ the following differential equations exist:
\begin{eqnarray}
\label{diff-I11}
\frac{d I_{10}}{d x} &=& \frac{2D-5+x^2(3-D)}{x(1+x)(1-x)}I_{10} +\frac{6-3D}{4x(1+x)(1-x)}I_{11}+ F_1\ , \\
\frac{d I_{11}}{d x} &=& \frac{4x(D-2)}{(1+x)(1-x)}I_{10} +\frac{3x(2-D)}{(1+x)(1-x)}I_{11}+ F_2
\end{eqnarray}
with
\begin{eqnarray}
F_1 &=& \frac{D-2}{4x(1+x)(1-x)}(I_1+2I_3)\ ,\\
F_2 &=& \frac{x(D-2)}{(1-x)(1+x)}I_1+\frac{2D-4}{x(1+x)(1-x)}I_3\ .
\end{eqnarray}
In solving the above differential equations, we determine the bases in a similar way as in the case for $I_8$ and $I_9$, and then the differential equations can also be solved iteratively. Note here the logarithmic functions may appear when transforming the differential equations into the form of (\ref{norm}), however the bases can transform the above differential equations into a Strictly Triangular Matrix when setting $D=4$. The bases we find are
\begin{eqnarray}
g_{10} = \frac{x^2+1}{x^3}I_{10}-\frac{I_{11}}{2x^3}\ ,\ \; g_{11} = \frac{I_{10}}{(x^2-1)^2}\ .
\end{eqnarray}
Then the analytic form of $I_{10}$ up to order $\epsilon^2$ is obtained by solving the differential equations:
\begin{eqnarray}
I_{10} &=& \frac{1+2x^2}{2\epsilon^2}+\frac{1}{\epsilon} \big(\frac{5}{4}+3x^2-4x^2H_0(x)\big)+\frac{11}{8}+6x^2
+\frac{(5-6x^2+4x^4)\pi^2}{12}\nonumber \\
& &-14x^2H_0(x)+4x^2(2+x^2)H_{0,0}(x)+ 2(x^2-1)^2H_{-,0}(x)+\epsilon\big[-\frac{55}{16}\nonumber\\
& &+\frac{15x^2}{2}+(\frac{25}{24}+2x-\frac{5x^2}{2}+2x^3
+\frac{5x^4}{6})\pi^2-\frac{x^2(111+2\pi^2)H_0(x)}{3}\nonumber\\
& &+(x^2-1)^2\pi^2(H_{-}(x)-H_{+}(x))+2x^2(12+5x^2)H_{0,0}(x)\nonumber\\
& &+4x(1+x^2)H_{+,0}(x)+(5-18x^2+5x^4) H_{-,0}(x)-8x^2(2+3x^2)H_{0,0,0}(x)\nonumber\\
& &+2(x^2-1)^2(3H_{-,-,0}(x)-H_{+,+,0}(x))+ \frac{11-26x^2+12x^4}{3}\zeta(3)\big] \nonumber\\
& &+\epsilon^2\big[-\frac{949}{32}-\frac{21x^2}{4}+ (55+624x-456x^2+624x^3+44x^4)\frac{\pi^2}{48}\nonumber\\
& &-16x(1+x^2)\pi^2\ln(2)-x\big(48(1+x^2)\pi^2+ x(525+14\pi^2-16\zeta(3))\big)\frac{H_0(x)}{6}\nonumber\\
& &+(1+x)^2\big(16\ln(2)-5+x(14-32\ln(2))+ x^2(16\ln(2)-5)\big)\frac{\pi^2H_+(x)}{2}\nonumber\\
& &+(1+x)^2\big((5-14x+5x^2)\pi^2+24(1-x)^2\zeta(3)\big) \frac{H_-(x)}{2}+(\frac{55}{6}-38x^2+10x^4)\zeta(3)\nonumber\\
& &+(303-578x^2+296x^4)\frac{\pi^4}{720}+ \frac{x^2(4(36+\pi^2)+x^2(33+2\pi^2))H_{0,0}(x)}{3}\nonumber\\
& &+\frac{(1+x^4)(33+2\pi^2)-2x^2(189+2\pi^2)} {6}H_{-,0}(x)+(26x(1+x^2)\nonumber\\
& &+4(x^2-1)^2\pi^2)H_{+,0}(x)+3(x^2-1)^2 \pi^2(H_{-,-}(x)-H_{-,+}(x))\nonumber\\
& &+(x^2-1)^2\pi^2(H_{+,-}(x)-H_{+,+}(x))- 4x^2(8+15x^2)H_{0,0,0}(x)\nonumber\\
& &-16x(1+x^2)H_{0,+,0}(x)+(5-18x^2+5x^4) (3H_{-,-,0}(x)-H_{+,+,0}(x))\nonumber\\
& &+4x(1+x^2)(3H_{+,-,0}(x)-H_{-,+,0}(x))+ 16x^2(2+7x^2)H_{0,0,0,0}(x)\nonumber\\
& &-16(x^2-1)^2H_{-,0,0,0}(x)+(x^2-1)^2 (18H_{-,-,-,0}(x)-6H_{-,+,+,0}(x)\nonumber\\
& &+8H_{+,0,+,0}(x)+2H_{+,-,+,0}(x)-6H_{+,+,-,0}(x))\big]\ .
\end{eqnarray}
The $I_{11}$ hence can be simply obtained from Eq.(\ref{diff-I11}) and will not be shown here.

The master integral $I_{12}$ can be obtained from integral $I_{10}$ by exchanging $m_b$ with $m_c$ in the norm and replacing $x$ by $1/x$, while the integral $I_{13}$ achieves in a similar way as $I_{11}$. All the analytic results we calculated have been numerically checked with {\bf FIESTA} \cite{fiesta}. The integral $I_{12}$ was given in \cite{remid}, while we find a misprint there that the first and second terms of (5.20) on page 397 in Ref.\cite{remid} should be multiplied by a factor of 1/8, otherwise taking the $x \rightarrow 1$ limit one can not get the $x=1$ result in the same paper.


After the calculation of master integrals in Fig.1, one can then start the renormalizeation procedure to remove the divergences encountered. The renormalization is performed by subtracting the one-loop sub-divergencies and the two-loop overall divergencies. The quark wave functions are renormalized in the on-shell scheme, while the strong coupling constant $\alpha_s$ is renormalized in the $\overline{\text{MS}}$ scheme. The NNLO renormalization for $B_c$ leptonic decays is similar to what shown in Ref. \cite{bonc}. After performing the one-loop mass and coupling constant renormalization, and two-loop wave function renormalization, the decay width can be expressed as
\begin{eqnarray}
\Gamma &=& Z_{2,b}^{\frac{1}{2}}Z_{2,c}^{\frac{1}{2}} \Gamma_{bare}(\alpha_{s0})\ ,
\end{eqnarray}
which can be expanded perturbatively as:
\begin{eqnarray}
\Gamma &=& \Gamma^{0l}+a\Gamma^{1l}+a^2\Gamma^{2l}+\mathcal{O}(a^3)\ , \\
\Gamma_{bare} &=& \Gamma^{0l}+a_0 \Gamma^{1l}_{bare}+a_0^2 \Gamma^{2l}_{bare} + \mathcal{O}(a_0^3)\ , \\
Z_{2,b} &=& 1+ a_0\delta Z_{2,b}^{1l}+ a_0^2\delta Z_{2,b}^{2l}+\mathcal{O}(a_0^3)\ , \\
Z_{2,c} &=& 1+ a_0\delta Z_{2,c}^{1l}+ a_0^2\delta Z_{2,c}^{2l}+\mathcal{O}(a_0^3)\ , \\
a_0 &=& a(1+ a\delta Z_{\alpha_s}^{1l}+ a^2\delta Z_{\alpha_s}^{2l}+\mathcal{O}(a^3))\ .
\end{eqnarray}
Here $a$ and $a_0$ denote respectively $\frac{\alpha_s}{\pi}$ and $\frac{\alpha_{s0}}{\pi}$. $\alpha_s$ is renormalized coupling constant and $\alpha_{s0}$ is bare coupling constant. The two-loop renormalized amplitude can be then reexpressed as:
\begin{eqnarray}
\Gamma^{2l} &=& \Gamma^{2l}_{bare} + \big\{\frac{1}{2}\delta Z_{2,b}^{2l}+ \frac{1}{2}\delta Z_{2,c}^{2l}+\frac{1}{2}\delta Z_{\alpha_s}^{1l}(\delta Z_{2,b}^{1l}+\delta Z_{2,c}^{1l})-\frac{1}{8}(\delta (Z_{2,b}^{1l})^2+(\delta Z_{2,b}^{1l})^2)\big\}\Gamma^{0l}\nonumber \\
& &+\big\{\frac{1}{2}\delta Z_{2,b}^{1l}+\frac{1}{2}\delta Z_{2,c}^{1l}+\delta Z_{\alpha_s}^{1l}\big\}\Gamma^{1l}_{bare}\ .
\end{eqnarray}
\begin{figure}[h,t,m,u]
\begin{center}
\includegraphics[scale=0.4]{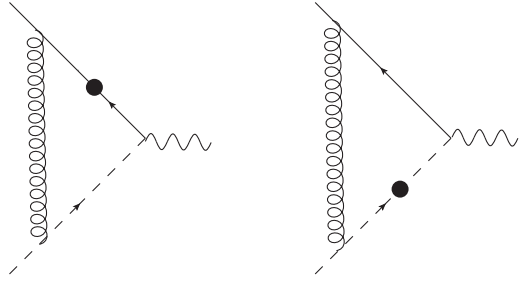}
\caption{Mass-renormalization counter-diagrams. \label{fig3}}
\end{center}
\end{figure}

For mass renormalization, there are two diagrams shown in Fig.3 to account for the one-loop mass renormalization. The one-loop coupling and mass renormalization constants as well as two-loop wave functions renormalization constants can be obtained from \cite{l2z2,z22}.


After the renormalization procedure, we can classify the two-loop coefficient $co_2$  in color factors as:
\begin{eqnarray}
co_2 &=& (C_F s_F+C_A s_A+N_L T_F s_L+ T_F s_H)C_F+\frac{\beta_0}{4}co_1\ln\frac{\mu^2}{m_b^2}\
\end{eqnarray}
with
\begin{eqnarray}
co_1 &=& \frac{3}{4}\big(\frac{x-1}{1+x}\ln x-2\big)C_F\
\end{eqnarray}
being the one-loop coefficient, $N_L$ the number of light quarks. The coefficients $s_F, s_A, s_L$ and $s_H$ read as:
\begin{eqnarray}
s_F &=& -\frac{(1+6x+x^2)\pi^2}{8(1+x)^2}\big(\frac{1} {\epsilon}+4\ln\frac{\mu_f}{m_b}\big)
+\frac{29}{16}-\frac{(85+151x+79x^2+x^3) \pi^2}{48(1+x)^2}\nonumber \\
& &+\pi^2\ln2-\frac{6x\zeta(3)}{(1+x)^2}+\frac{123+32\pi^2+200\pi^2x+
(16\pi^2-123)x^2+8\pi^2x^3}{96(1+x)^2}H_0(x)\nonumber \\
& &+\frac{(1-x)^2(1+x^2)\pi^2}{16x(1+x)^2}(H_-(x)-H_+(x))
+\frac{9+22x-15x^2-4x^3}{16(1+x)^2}H_{0,0}(x)\nonumber \\
& &+\frac{x-1}{2(1+x)}H_{+,0}(x)+\frac{1+5x-5x^2-x^3}{8x(1+x)}H_{-,0}(x)
-\frac{1+6x+x^2}{4(1+x)^2}H_{+,0,0}(x)\nonumber \\
& &+\frac{1-3x-4x^2-3x^3+x^4}{4x(1+x)^2} (H_{-,0,0}(x)-H_{0,-,0}(x))+\frac{1+6x+x^2}{4(1+x)^2}H_{0,+,0}(x)\ ,
\end{eqnarray}
\begin{eqnarray}
s_A &=& -\frac{\pi^2}{8}(\frac{1}{\epsilon}+4\ln\frac{\mu_f}{m_b})- \frac{17}{48}+\frac{(35-x)\pi^2}{96}
-\frac{9}{4}\zeta(3)-\frac{\pi^2\ln 2}{2}\nonumber \\
& &+\frac{115+52\pi^2+(149+48\pi^2)x-4\pi^2x^2}{96(1+x)}H_0(x)+ \frac{(1-10x+x^2)\pi^2}{32x}(H_+(x)\nonumber \\
& &-H_-(x))+\frac{11-9x+2x^2}{8(1+x)}H_{0,0}(x)+ \frac{1-x^2}{16x}(H_{+,0}(x)-2H_{-,0}(x))\nonumber \\
& &+\frac{1}{4}(H_{0,+,0}(x)-H_{+,0,0}(x))+ \frac{1-8x+x^2}{8x}(H_{0,-,0}(x)-H_{-,0,0}(x))\ ,
\end{eqnarray}
\begin{eqnarray}
s_L &=& \frac{1}{12}-\frac{11+13x}{24(1+x)}H_0(x)+ \frac{x-1}{2(1+x)}H_{0,0}(x)\ ,
\end{eqnarray}
and
\begin{eqnarray}
s_H &&= \frac{35}{12}+\frac{9}{8} (x+\frac{1}{x})+(x^2+\frac{1}{x^2})
-\frac{(19+42x+66x^3+13x^4-12x^5)\pi^2}{96(1+x)}\nonumber \\
& &-\frac{15+36x+13x^2+35x^3-36x^4-15x^5} {24x^2(1+x)}H_0(x)+\frac{-1+x+5x^4+3x^5}{2(1+x)}H_{0,0}(x)\nonumber \\
& &+\frac{11+22x+5x^3-5x^4-22x^6-11x^7}{16x^3(1+x)}H_{+,0}(x)\nonumber \\
&&+\frac{-3-5x+2x^4-2x^5+5x^8+3x^9}{4x^4(1+x)}H_{-,0}(x)\ .
\end{eqnarray}

It is notable that after taking the renomalization procedure in above, the result is still divergent, which can be attributed to the anomalous dimension of NRQCD current \cite{js2lo}. In $\overline{\text{MS}}$ scheme, the anomalous dimension of NRQCD current first arises at two-loop order as noticed and defined in Refs.\cite{js2lo,bc22}. For our case, the anomalous dimension reads:
\beq
\gamma_{J,NRQCD}=\frac{\text{d}\ln Z_{J,NRQCD}}{\text{d}\ln\mu}= (-\frac{1+6x+x^2}{2(1+x)^2}C_F^2-\frac{C_F C_A}{2})\alpha_s^2+\mathcal{O}(\alpha_s^3)\ .
\eeq

In the literature, the two-loop QCD corrections to $B_c$ leptonic decays, the coefficient $co_2$, was once investigated under the condition of small parameter $x=\frac{m_c}{m_b}$ expansion up to the second order $(\frac{m_c}{m_b})^2$ \cite{bc22}. For $s_F, s_A$ and $s_H$ we expand our complete analytic results in $x$ to the second order and find an agreement with Ref. \cite{bc22}, while for $s_L$, we find there is a misprint of redundant term $\frac{5\pi^2}{144}$. We notice that in the attachment files of \cite{bc22} the renormalization constant $Z_2$ disagrees with the one given in \cite{l2z2,z22}. We have applied our calculation procedure to the study of two-loop corrections to $J/\psi$ and $\Upsilon$ leptonic decays, and found full agreement with Refs. \cite{js2lo,czme}, analytically.


To evaluate numerically the $B_c$ leptonic decay rates to the next-to-next-to-leading order degree of accuracy, we take the following input parameters \cite{PDG,eichten,Bodwin:2007zf}:
\begin{eqnarray}
m_c &=& 1.5\ \text{GeV}\ ,\ m_b= 4.8\ \text{GeV}\ ,\ G_F =1.16637\times 10^{-5}\ \text{GeV}^{-2}\ ,\nonumber\\
|V_{cb}| &=& 0.0406\ ,\ f_{B_c}^{NR}= 0.499\ \text{GeV}\ ,\ N_L=3\ ,\ \Lambda_{QCD} = 0.214\ \text{GeV}\ ,\\
m_{\mu} &=& 0.106\ \text{GeV}\ ,\ m_{\tau}=1.777\ \text{GeV}\ ,\ \tau(B_c)=0.509\ \text{ps}\nonumber \ .
\end{eqnarray}
Here, the non-relativistic decay constant $f_{B_c}^{NR}$, defined as $f_{B_c}^{NR} = -i \langle0|\chi_b^\dagger \psi_c|B_c\rangle/M_{B_c}$ \cite{js2lo}, is obtained by potential model evaluation \cite{eichten}, the pole charm- and bottom-quark masses are adopted \cite{Bodwin:2007zf}, and $\tau(B_c)$ is the $B_c$ life time. For numerical estimates we take factorization scale $\mu_f=1$ GeV to separate perturbative and nonperturbative domains and the renormalization scale $\mu$ is set to be at bottom quark mass. In the end, the decay constant in $\overline{\text{MS}}$ factorization scheme can be expressed as:
\begin{eqnarray}
f_{B_c} &=& (1-1.39(\frac{\alpha_s(m_b)}{\pi})-23.7 (\frac{\alpha_s(m_b)}{\pi})^2)f_{B_c}^{NR}\nonumber\\
&=&(1-0.094-0.108)f_{B_c}^{NR}=0.798f_{B_c}^{NR}\ .
\label{decay-constant}
\end{eqnarray}
With which, we can immediately obtain the branching ratios of $B_c$ leptonic decays at NNLO accuracy, i.e.,
\begin{eqnarray}
 Br(B_c\rightarrow \tau^+\nu_{\tau})\approx1.8\times 10^{-2}\ ,\ Br(B_c\rightarrow \mu^+\nu_{\mu})\approx 7.6\times 10^{-5} \ .
\end{eqnarray}
The branching fraction of $B_c$ decay to positron and neutrino is much smaller than the numbers in above as expected.

It is notable that Eq.(\ref{decay-constant}) indicates that the NNLO corrections are greater than the NLO corrections, similar as the case of quarkonium leptonic decays. However, the recent result on 3-loop corrections for $\Upsilon$ leptonic decays turns out that the N$^3$LO contribution is small and the renormalization scale dependence is greatly reduced \cite{Beneke:2014qea}.


In summary, we calculated analytically the two-loop QCD corrections to $B_c$ meson leptonic decays in the framework of NRQCD. All the master integrals were achieved analytically by means of Mellin-Barnes integral or Differential Equations. We expanded our  analytic results in parameter  $x=\frac{m_c}{m_b}$ to the second order and found a partial agreement with the results in previous calculation. To confirm our calculation, we restudied the NNLO QCD corrections to heavy quarkonium leptonic decays and can fully reproduce those results in the literature. With proper inputs, we numerically computed the $B_c$ leptonic decay widths to the full next-to-next-to-leading order degree of accuracy, and found the the NNLO corrections are remarkable. This calculation may
be helpful as well to the precision measurement of Cabibbo-Kobayashi-Maskawa(CKM) matrix element $|V_{cb}|$ when the $B_c$ meson decay constant is well determined by lattice QCD calculation.

\vspace{.7cm}
{\bf Acknowledgments}

This work was supported in part by the Ministry of Science and Technology of the People's Republic of China (2015CB856703), and by the National Natural Science Foundation of China(NSFC) under the grants 11175249 and 11375200.




\end{document}